\documentclass[trackchanges]{aastex7}

\usepackage{tabularx}
\usepackage{longtable}
\usepackage{multirow}
\usepackage{booktabs}
\usepackage{amsmath}

\usepackage[utf8]{inputenc}
\DeclareUnicodeCharacter{2212}{-}

\definecolor{DarkGreen}{rgb}{0.0,0.4,0.0}  
\newcommand{\B}[1]{{\color{blue}{#1}}}
\newcommand{\R}[1]{{\color{red}{#1}}}
\newcommand{\G}[1]{{\color{DarkGreen}{#1}}}
\newcommand{\M}[1]{{\color{magenta}{#1}}}
\newcommand{\C}[1]{{\color{cyan}{#1}}}
\usepackage[normalem]{ulem}
\usepackage{soul}
\usepackage{cancel}

\begin{document}


\title{Investigation on the Relation between Active Regions' Compliance with Empirical Laws and Flare Productivity}

\author[orcid=0009-0008-3359-8092, gname=Jinhui, sname='Pan']{Jinhui Pan}
\affiliation{CAS Key Laboratory of Geospace Environment, Department of Geophysics and Planetary Sciences, University of Science and Technology of China, Hefei 230026, People’s Republic of China}
\email{panjh@mail.ustc.edu.cn}  

\author[orcid=0000-0003-4618-4979, gname=Rui, sname='Liu']{Rui Liu} 
\affiliation{CAS Key Laboratory of Geospace Environment, Department of Geophysics and Planetary Sciences, University of Science and Technology of China, Hefei 230026, People’s Republic of China}
\affiliation{Mengcheng National Geophysical Observatory, University of Science and Technology of China, Hefei 230026, People’s Republic of China}
\email[show]{rliu@ustc.edu.cn}

\author{Jiangtao Su}
\affiliation{State Key Laboratory of Solar Activity and Space Weather, National Astronomical Observatories, Chinese Academy of Sciences, Beijing 100012, People’s Republic of China}
\affiliation{University of Chinese Academy of Sciences, Beijing 100049, People’s Republic of China}
\email{sjt@bao.ac.cn}

\author[orcid=0000-0001-5002-0577, gname=Jie, sname='Jiang']{Jie Jiang}
\affiliation{School of Space and Earth Sciences, Beihang University, Beijing, People’s Republic of China}
\affiliation{Key Laboratory of Solar Activity, National Astronomical Observatories, Chinese Academy of Sciences, Beijing 100012, People's Republic of China}
\email{jiejiang@buaa.edu.cn}
\begin{abstract}
It remains evasive whether solar active regions (ARs) obeying or violating Hale's polarity law, Joy's tilt law, and the hemispheric helicity rule (HHR) differ in flare productivity. Here we conduct a comprehensive statistical analysis of ARs during the Solar Cycle 24 and the ascending phase of Cycle 25. ARs are automatically detected from full-disk line-of-sight magnetograms acquired by the Michelson Doppler Imager (MDI) and the Helioseismic and Magnetic Imager (HMI). We calculate tilt angles via flux-weighted polarity centroids, estimate magnetic twist by the force-free parameter $\alpha_{\mathrm{best}}$ from HMI vector magnetograms, and measure flare productivity using the flare index (FI) built from GOES C-class-and-above events. Our results substantiate that the majority of ARs follow the aforementioned three empirical laws. The compliance rate tends to be higher for ARs emerging at higher latitudes or having larger centroid distance, while total unsigned magnetic flux exerts limited influence, with a clear positive correlation only for Hale's law. Overall, FI shows no significant discrepancies across different compliance groups, except that Cycle 24 ARs that satisfy Hale's and Joy's laws but violate the HHR exhibit higher FI than other groups. We also identify empirical thresholds for centroid distance and total unsigned flux, above which the median FI of binned ARs becomes nonzero. Combining the flux and distance thresholds effectively separates flare-productive from flare-quiet ARs. We hence conclude that the flare productivity of ARs is not dependent on the compliance with the empirical laws, but more closely associated with sufficiently large and strong magnetic systems.  

\end{abstract}


\section{Introduction}     \label{S-Introduction} 
Solar active regions (ARs) are the most predominant manifestation of the large-scale magnetic field across the solar surface \cite[]{van2015evolution,2019LRSPToriumi}. Due to the substantial magnetic free energy contained in the ARs, they always serve as the primary sources of the solar flares and coronal mass ejections, the most intense eruptions in the heliosphere. These eruptive phenomena can drive severe space weather events, e.g., disrupting satellite communications, navigation systems, and power grids on Earth \cite[]{Temmer2021}. Therefore, understanding the relationship between the properties of ARs and solar eruptions is of paramount importance.

As \cite{1994ApJParker} suggested, sunspots and ARs are formed when buoyant magnetic flux tubes in the convection zone break through the solar surface into the atmosphere. During this transit through the convective regions, the dynamic evolution of the rising flux tubes is governed by the interaction between magnetic buoyancy, convective turbulence, and the solar rotation. Consequently, the emerging ARs do not appear randomly but exhibit statistical patterns that reflect the inner work of the global solar magnetic field. Empirically, bipolar ARs in the same hemisphere tend to have the same leading magnetic polarity, whereas those in the opposite hemisphere have opposite leading polarities. This hemispheric polarity pattern reverses from one solar cycle to the next, and is known as Hale's law or the Hale--Nicholson law \citep{1919ApJHale}. Moreover, the leading polarities are closer to the solar equator than the following ones \citep{1919ApJHale}. The tilt angle, which measures the inclination of the AR polarity axis relative to the solar equator, generally increases with latitude which is known as Joy's law \citep{1988SciZirin}. The physical origins of these empirical laws are closely tied to the dynamics of the rising flux tubes. This flux-emergence picture has been investigated through the thin flux tube model and MHD simulations. In the thin flux tube approximation, the rising tube is treated as a slender magnetic structure subject to magnetic buoyancy, magnetic tension, and the Coriolis force. This model has been used to explain the Joy's law tilt angle, the latitude dependence of emergence, and the asymmetry between the leading and following polarities \citep{1993A&ADsilva, Fan1993, Fan1994, 1995ApJCal}. In this framework, the angle of tilt described by Joy's law is generally attributed to the Coriolis force acting on the rising flux tubes during their ascent through the convection zone, which tilts the tube axis relative to the east-west direction \citep{2011ApJWeber, 2015SoPh..290.1295W}. More complete 2D and 3D MHD simulations further resolve the cross section and magnetic twist of the tube, showing that the twist of magnetic field lines is important for maintaining the coherence of rising flux tubes and for producing AR-like emergence properties \citep{Fan1998, Fan2008}. Alternatively, Joy’s Law may be related to the conservation of magnetic helicity in a rising flux tube as it breaks through the surface \cite[e.g.,][]{1997ApJLoncope,1998ApJLongcope}. 

However, these laws are statistical rather than absolute. Flux-emergence simulations that include convective motions show that turbulent convection can modulate the rise, deformation, and final tilt of magnetic flux tubes, especially when the tube field is not substantially super-equipartition \citep{Fanyuhong2003, 2009ApJJouve, 2011ApJWeber, 2013SoPhWeber}. During their ascent, flux tubes are subject to the buffeting of convective turbulence, which competes with the magnetic tension of the flux tubes \cite[]{1995ApJFisher}. Theoretical models and simulations suggest that the compliance with these laws is strongly dependent on the physical properties of ARs, particularly their magnetic flux and the separation of the polarities \cite[]{1993A&ADsilva, 2018ApJ...867...89L, 2024ApJ...976...20W}. Stronger flux tubes possess sufficient magnetic tension to resist convective disruption, thereby preserving the systematic tilt angles and polarity signatures imprinted by the solar dynamo and Coriolis force. In contrast, weaker flux tubes are more susceptible to the turbulent flows, leading to a randomized distribution of tilt angles and violation of the empirical laws \citep{1989SoPhWang,Stenflo&Kosovichev2012,Fanyuhong2003, 2009ApJJouve}. The simulations of global convection and dynamo action provide a complementary picture in which buoyant magnetic loops and AR-like emerging flux systems are generated self-consistently by turbulent convection and large-scale dynamo action, rather than isolated flux tubes. These structures generated by dynamo and convection action can exhibit statistical properties relevant to the empirical laws. \citep{Nelson2014,2014ApJFanFang}.

Besides the geometric orientation and the magnetic polarities described by Joy's law and Hale's law, the internal magnetic topology of ARs exhibits a hemispheric preference known as the Hemispheric Helicity Rule (HHR). Magnetic helicity quantifies the complexity of the magnetic field in terms of twists, kinks, and the linkages of field lines \cite[]{1984Berge}. Extensive observations have established that ARs in the northern (southern) hemisphere predominantly exhibit negative (positive) helicity corresponding to a left-handed (right-handed) twist, ranging across the solar atmosphere from photospheric magnetic fields to coronal sigmoids \cite[]{1987GeoRLBieber, 1998ApJBao, 1999GeoRLCanfield, 2014SoPhSavcheva, 2019LRSPToriumi}. The physical origin of the HHR is related to the dynamics of flux tubes traversing the convection zone. Several mechanisms have been proposed to explain the generation of this empirical hemispheric bias, such as the $\Sigma$-effect \cite[]{1998ApJLongcope}, which suggests that helical turbulence in the convection zone imparts a systematical twist to rising flux tubes. Additionally, the Coriolis force acting on the expanding material within a rising tube can generate twist consistent with the HHR \cite[]{1985ApJGlatzmaier, 1997ApJLoncope, 2013ApJWYM}, while differential rotation may further shear the magnetic footpoints on the surface \cite[]{1990SoPhSeehafer, 2000JGRBerger, 2002SoPhDemoulin}. Consequently, the compliance of an AR with the HHR could act as a tracer for the dynamo processes and the turbulent interactions occurring deep in the solar interior. 

Similarly, the adherence to the HHR is statistical rather than absolute. Some observational studies indicate that 60--70\% of ARs follow HHR, while a significant fraction violates it \cite[]{2014ApJLiu, 2020ApJPark}. These violations are of particular interest because magnetic twist is a manifestation of current-carrying non-potential magnetic fields, which store the free magnetic energy necessary for solar eruptions. \cite{2005SoPhTian} highlighted that complex $\delta$-type ARs often exhibit inconsistencies between their magnetic twist and writhe, which are quantified by the magnetic helicity and the tilt angles of the ARs, respectively. These authors concluded that ARs violating Hale's law but obeying the HHR have a stronger tendency to produce X-class flares. Recently \cite{2021ApJPark} found that ARs with low HHR compliance can be associated with enhanced flaring activity, suggesting that the injection of anomalous helicity or the interaction between systems of opposite chirality may be a critical driver for solar eruptions. On the other hand, extensive statistical investigations comparing flare-active and flare-quiet regions have emphasized the consensus that solar eruptions are preferentially triggered in ARs possessing specific magnetic signatures, such as complex configurations ($\delta$-type ARs), high degree of magnetic non-potentiality, and intense, dynamic evolution in the vicinity of polarity inversion lines \cite[]{1987SoPhZirin, 2007ApJGeorgoulis, ChenAQ2012, 2015SCPMAWand, 2019LRSPToriumi}.

While it has been established for decades that the magnetic complexity and local properties of ARs are closely related to the flare productivity, the relationship between the compliance with empirical laws and flare productivity remains an open question. In this study, we characterize the statistical distribution of ARs with respect to these empirical laws using a comprehensive data set spanning Solar Cycles 24 and 25 (hereafter SC24 and SC25 respectively). We also quantitatively investigate the correlation between the empirical law compliance and flare productivity with rigorous statistical tests. The rest of the paper is organized as follows: Section~\ref{S-DATA} describes the data set construction and the methodology for parameter extraction; Section~\ref{S-sre} outlines the statistical method and presents the statistical results; Section~\ref{S-discussion} discusses and summarize the results.

 
\section{Data and Method} 
      \label{S-DATA}      
\begin{figure} 
\centerline{\includegraphics[width=1\textwidth,clip=]{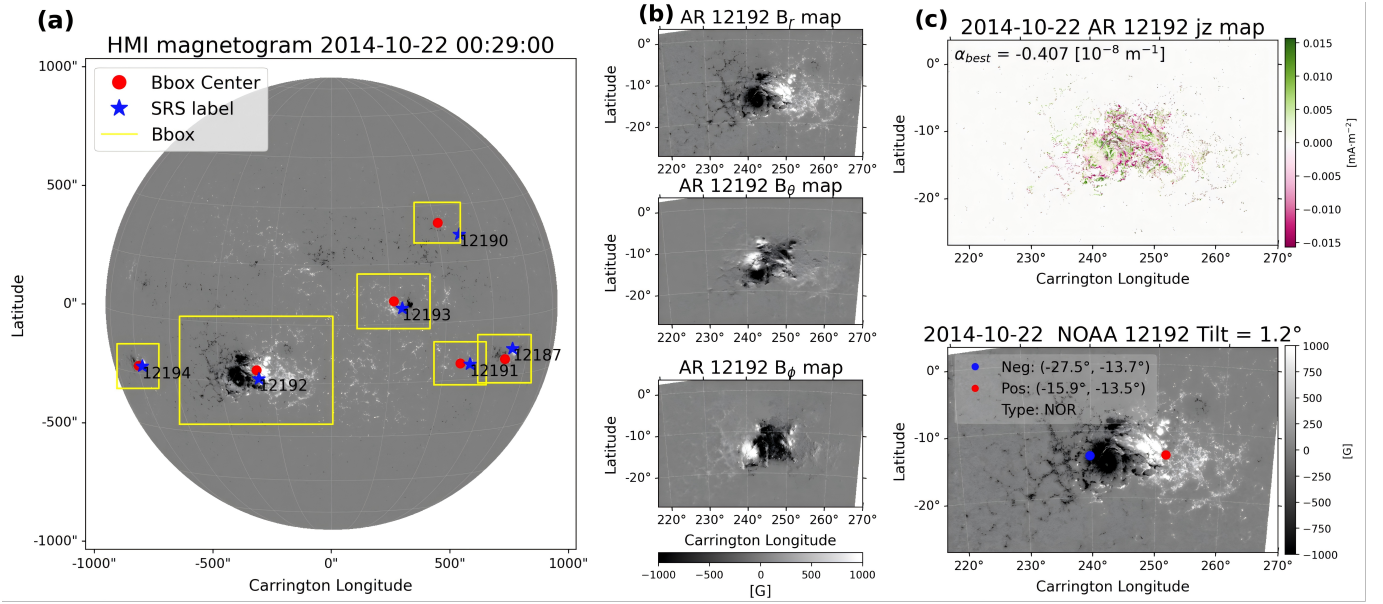}}
\small
        \caption{Data processing pipeline. (a) SARD detection results on the full-disk LoS magnetogram on 2014 October 22, which are calibrated with the AR information from the NOAA SRS catalog. The detected ARs are indicated by yellow bounding boxes (Bbox)}, whose centers are marked by red dots, in comparison with the AR locations in the NOAA SRS catalog, as marked by blue asterisks. (b) (B$_r$, B$_\theta$, B$_\phi$) maps of AR 12192 in the Heliographic Carrington coordinates, extracted and converted from the HMI full-disk vector magnetograms according to the SARD detection results. Magnetic field values are saturated at $\pm1000$~G. (c) Top: Map of j$_z$ for AR 12192. The derived $\alpha_{\mathrm{best}}$ is negative, as annotated in the upper left, indicating that this southern-hemisphere AR is inconsistent with the HHR. Bottom: B$_r$ map of AR 12192, where the magnetic flux weighted centroids of the positive and negative polarities are marked by blue and red dots, respectively. With a tilt angle of 1.2$^\circ$, this southern-hemisphere AR is consistent with both Hale’s polarity law and Joy’s law, hence classified as NOR.
\label{fig-pipeline}
\end{figure}

In this study, we employ the Solar Active Region Detection (SARD) model \cite[]{2025panmodel} to automatically extract ARs from the solar full-disk line-of-sight (LoS) magnetograms. SARD is a deep learning based object detector, built on the YOLOv8 architecture and trained on an HMI LoS magnetogram data set spanning from 2010 to 2019 \cite[]{2012HMI,2025pandataset}. This model demonstrates a better performance than the traditional detection methods based on image processing techniques \citep[see details in][]{2025panmodel}. 
To improve the association between model detections and NOAA AR labels, we cross-match SARD detection results with AR locations reported in the NOAA Solar Region Summary (SRS) issued by the NOAA Space Weather Prediction Center\footnote{\url{https://www.swpc.noaa.gov/products/solar-region-summary}} and retain only those detection results consistent with the SRS locations. An exemplary SARD detection result from the full-disk LoS magnetogram on 2014 October 2 is shown in Figure~\ref{fig-pipeline}a. The yellow bounding boxes denote the retained SARD detection results. The red dots mark the centers of the detected boxes, while the blue asterisks indicate the AR locations reported in the NOAA SRS catalog.

Our LoS magnetogram data set consists of (i) SOHO/MDI full-disk LoS magnetograms from 2008 December 1 to 2010 April 30 (\texttt{mdi.fd\_M\_96m\_lev182}), and (ii) SDO/HMI full-disk LoS magnetograms from 2010 May 1 to 2025 December 31 (\texttt{hmi.M\_720s}). The vector magnetograms are derived from the HMI full-disk vector magnetogram series \texttt{hmi.B\_720s}, sampled once per day at 00:36:00 TAI from 2010 May 1 to 2025 December 31. The \texttt{hmi.B\_720s} products provide the magnetic field strength, inclination, and azimuth, together with the corresponding 180$^\circ$ disambiguation information. Alternatively, 
the HMI pipeline detects ARs in photospheric LoS magnetograms and intensity images, and assigns an HMI Active Region Patch (HARP) number to each AR crossing the face of the solar disk, but a single HARP number may correspond to multiple NOAA AR labels. To ensure the consistency with the NOAA AR labeling, we used the SARD detection results for convenience.

In order to obtain the magnetic field components for each AR, we follow the coordinate remapping and vector magnetic field components transformation procedure adopted in the HMI Space Weather Helioseismic and Magnetic Imager AR Patches (SHARP) pipeline \cite[]{2013sharptrans,2014SHARP}. Specifically, we first convert the disambiguated field strength, inclination, and azimuth to the image-plane components $(B_\xi, B_{\eta}, B_{\zeta})$ on the helioprojective Cartesian basis $(\hat{\boldsymbol{e}}_\xi,\ \hat{\boldsymbol{e}}_\eta,\ \hat{\boldsymbol{e}}_\zeta)$, where $+\zeta$ denotes the line of sight. For each AR cutout, the image-plane components are remapped to a Cylindrical Equal Area \citep[CEA;][]{Calabretta&Greisen2002} grid centered at the center of the AR, and resampled with a uniform spacing of 0.03$^\circ$ in both directions. The newly sampled $(B_\xi, B_{\eta}, B_{\zeta})$ are then transformed to heliocentric spherical components $(B_r, B_\theta, B_\phi)$ using the transformation matrix given by Equation (1) in \cite{1990Gary}, as implemented in the HMI pipeline. In this convention, B$_{\theta}$ is defined to be positive when pointing toward south. Moreover, $(B_r, B_\theta, B_\phi)$ are identical to the heliographic components $(B^{h}_{z}, -B^{h}_{y}, B^{h}_{x})$ in \cite{1990Gary}. We work with ($B_{x}, -B_{y}, B_{z}$) in heliographic Cartesian coordinates for the subsequent calculations. The maps of $(B_r, B_\theta, B_\phi)$ of AR 12192 are shown in Figure~\ref{fig-pipeline}b, where the values of magnetic field are saturated at $\pm1000$~G. The white regions arise from edge effects introduced during the coordinate transformation of the AR cutouts extracted from the full-disk LoS magnetograms. These artifacts are confined to the periphery of the AR cutouts and therefore do not affect the subsequent calculations. 

\begin{figure} 
\centerline{\includegraphics[width=1\textwidth,clip=]{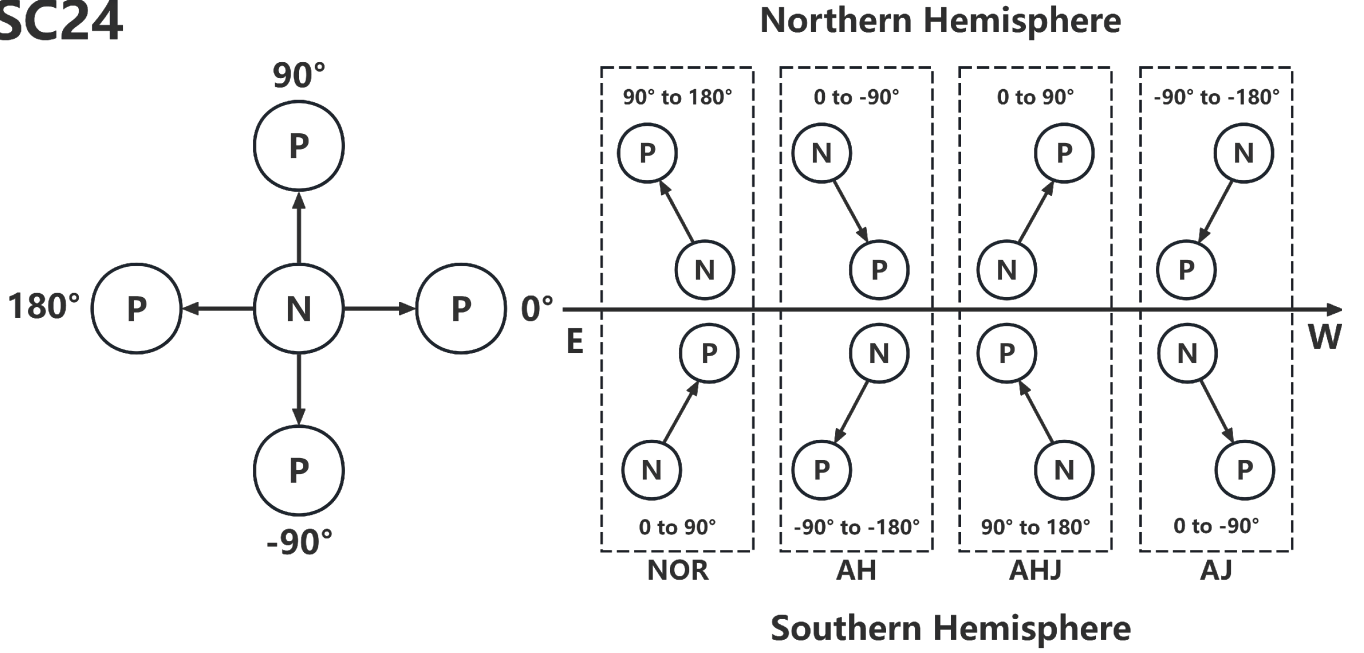}}
\small
        \caption{Definitions of the tilt angles for SC24. The left panel shows the adopted sign convention for the tilt angle, which spans $(-180^\circ,180^\circ]$. P denotes the positive polarity and N denotes the negative polarity. The right panel illustrates the classification of ARs according to their compliance with Hale's law and Joy's law. `NOR' (normal) denotes ARs that obey both Hale's law and Joy's law, `AH' (Anti-Hale) denotes ARs that violate Hale's law but obey Joy's law, `AJ' (Anti-Joy) denotes ARs that obey Hale's law but violate Joy's law, and `AHJ' (Anti-Hale-and-Joy) denotes ARs that violate both laws. The corresponding tilt-angle ranges for each class are annotated in the figure.} 
\label{fig-cartoon}
\end{figure}

In computing the AR tilt angle, we first identify the positive and negative polarities using the magnetic flux weighted centroids measured from the $B_r$ map of each AR (Figure~\ref{fig-pipeline}b). Let ($\lambda_+,~\phi_+$) and ($\lambda_-,~\phi_-$) denote the heliographic latitude and longitude of the positive and negative polarity centroids, respectively. We measure an AR's  orientation relative to the local east-west direction by the vector pointing from the negative to the positive centroid. The tilt angle $\gamma$ is then computed as follows \citep{1989SoPhWang},
\begin{equation}
    \tan \gamma = \frac{\Delta \lambda}{\Delta \phi \cos \bar{\lambda}},
\end{equation}
where $\Delta \lambda=\lambda_+-\lambda_-$, $\Delta \phi=\phi_+-\phi_-$, and $\bar{\lambda}=(\lambda_++\lambda_-)/2$. We not only calculate the values of tilt angles, but also assign the angle to the correct quadrant, so that the tilt angle varies between [−180$^\circ$, 180$^\circ$]. This method is applied uniformly in both northern and southern hemispheres, following \cite{1991Howard}. We consider only the pixels with $|B_r| > 100$~G, and estimate the error by adjusting the $B_r$ threshold from 100 to 150 G.
Cases in which the tilt angle changes sign or from acute to obtuse due to the different threshold selection are taken as uncertainties in determining the compliance rate for Joy's law.
The left panel of Figure~\ref{fig-cartoon} shows an illustration of the adopted sign convention for tilt angle $\gamma$. The right panel of Figure~\ref{fig-cartoon} presents an illustration of the classification of ARs based on their compliance with Hale's law and Joy's law in SC 24. If the tilt angle is positive obtuse (positive acute) in the northern (southern) hemisphere, the AR obeys both Hale's law and Joy's law, so it is labeled as `NOR' (normal). If the tilt angle is positive acute (positive obtuse) in the northern (southern) hemisphere, the AR violates both Hale's law and Joy's law and is labeled as `Anti-Hale-and-Joy' (AHJ). The `Anti-Hale' (AH) type consists of ARs that violate Hale's law but obey Joy's law which correspond to negative acute (negative obtuse) tilt angles in the northern (southern) hemisphere. The `Anti-Joy' (AJ) type consists of ARs that obey Hale's law but violate Joy's law which correspond to negative obtuse (negative acute) tilt angles in the northern (southern) hemisphere. It is noted that we treat AHJ as a separate class rather than merge it into AH or AJ.

The force-free parameter $\alpha$ is commonly used as a proxy for magnetic twist \cite[]{1990SoPhSeehafer,1995ApJPevtsov,1999Leka}. Under the force-free approximation, $\nabla\times\mathbf{B}=\alpha\mathbf{B}$. 
We obtain the ``best-fit'' $\alpha$ via a least-squares fitting over all pixels within the AR, i.e.,
\begin{equation}
    \alpha_\mathrm{best}=\frac{\sum B_z  (\nabla\times\mathbf{B})_z}{\sum {B_z^2}}.
\end{equation}
The summation is implemented over the entire AR, and only the pixels with $|B_x|$, $|B_y|$, and $|B_z|$ all exceeding 100~G are under consideration for robustness \cite[]{1999Leka, 2022A&ABaumgartner}. The top panel of Figure~\ref{fig-pipeline}c shows the $J_z$ map of AR 12192 on 2014 October 22, with the computed $\alpha_{\mathrm{best}}$ annotated in the top-left corner. This AR violates the HHR, because it is located in the southern hemisphere but the corresponding $\alpha_{\mathrm{best}}$ is negative. 

\begin{figure}[ht]
\centerline{\includegraphics[width=1\textwidth,clip=]{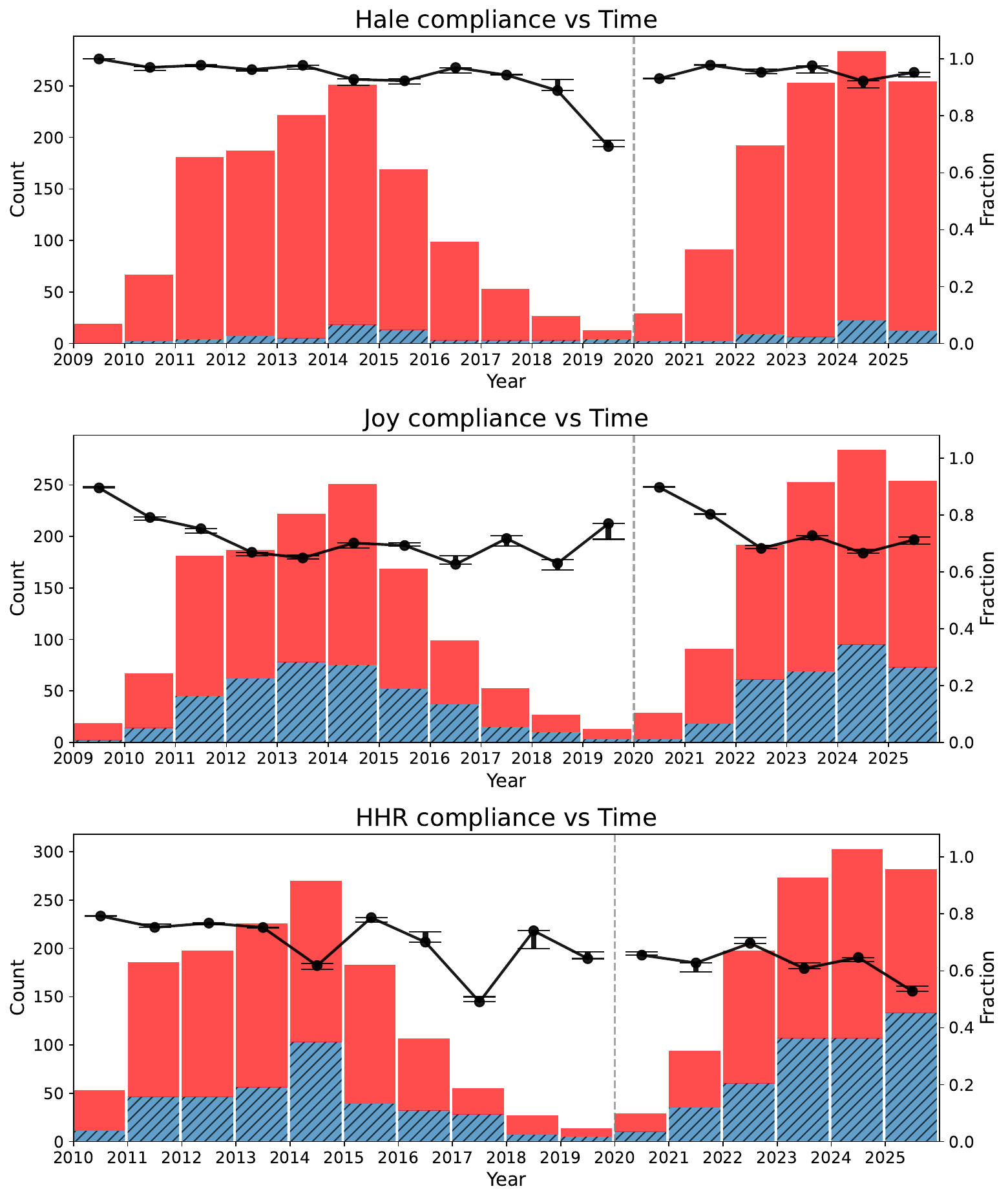}}
\small
        \caption{Frequency distribution of ARs that obey/violate Hale's law (top), Joy's law (middle), and the HHR (bottom), respectively. The red (blue) bars, as scaled by the left $y$-axis, represent the number of ARs that obey (violate) the empirical laws; black dots, as scaled by the right $y$-axis, show the yearly compliance rate. Error bars show the range of compliance rates obtained by varying the magnetic field threshold, the AR patch size, and the cycle boundary, as described in Section~\ref{S-DATA}.}
\label{fig-COMLIANCEVSTIME} 
\end{figure}

We further estimate the uncertainty of $\alpha_{\mathrm{best}}$ by (i) varying the field strength threshold from 100 to 150 G and (ii) enlarging the SARD detected AR patches by $2^\circ$. Cases in which $\alpha_{\mathrm{best}}$ changes sign under these variations are taken as uncertainties in determining the compliance rate for the HHR. For the binned compliance rates shown in Figures~\ref{fig-COMLIANCEVSLATITUDE} to \ref{fig-COMLIANCEVSFLUX}, the error bars represent the sensitivity of the compliance rates to the parameter choices, rather than standard binomial statistical errors. For each bin, we first compute the nominal compliance rate $f_0=n_{\rm obey}/n$ using the fiducial parameters. We then repeat the classification and recompute the compliance rate after varying the parameters including the magnetic field threshold, the AR patch size, and the cycle boundary between SC24 and SC25 (see \S\ref{SS-DISTRIBUTION2425}), to obtain a set of compliance rates $f$. The lower and upper bounds of the error bar then correspond to the minimum and maximum compliance rates obtained from these trials, respectively, i.e., $f_{\rm low}=\min f$ and $f_{\rm high}=\max f$. Since the trial results are not necessarily symmetric around the fiducial one, $f_0$ as shown by the black dot may not be in the center of the error bars spanning $[f_{\rm low},f_{\rm high}]$.

SARD is applied to each LoS magnetogram in the data set. To mitigate the projection effects, only the ARs that fall within [$-30^\circ,~30^\circ$] in longitude are retained. Because a given NOAA AR can be detected on multiple days during its disk passage, we assign each NOAA AR a single compliance label for Hale's law, Joy's law, and the HHR by the majority vote of all detection results on the same AR. To mitigate cases with severe polarity imbalance, we compute the flux imbalance ratio
\begin{equation}
R=\frac{|F_+ + F_-|}{|F_+|+|F_-|},
\end{equation}
where $F_+$ and $F_-$ are the total positive and negative magnetic flux in the AR. We flag a detection as unipolar if $R>0.5$ \cite[]{2017A&AVirtanen,2020SoPhYeates,2023ApJSWang}. An AR is finally labeled unipolar and excluded from subsequent analysis only if all detections during its disk passage are flagged as unipolar. For all the detections of each individual AR, we also compute its centroid distance and total unsigned flux, and retain the median values for statistics.


\section{Results} 
      \label{S-sre}      

\subsection{Distribution of the compliance of the laws across SC24 and SC25}
\label{SS-DISTRIBUTION2425}

After applying the detection, quality control, and parameter calculations described in Section~\ref{S-DATA}, we identify 2616 NOAA-labeled ARs in the interval from 2008 December 1 to 2025 December 31. Among the detection results, there are 156 unipolar ARs, which are excluded from the subsequent analysis. 45 ARs are detected from MDI full-disk LoS magnetograms (2008 December to 2010 April), and therefore missing the $\alpha_{\mathrm{best}}$ parameter. SC24 is commonly taken to span from the minimum in 2008 December to the minimum in 2019 December, with SC25 beginning in 2019 December \cite[]{2015LRSPHathaway}. Accordingly, we split the sample into SC24 (up to 2019 November 30) and SC25 (from 2019 December 1 onward). Since adjacent solar cycles may have an overlap of about three years \cite[]{1988Natur8Wilson, 1994SoPh7Hathaway, 2018ApJJiang}, we estimate the uncertainty associated with the cycle overlap by varying the specified cycle boundary from 2018 December 1 to 2020 December 1. For each trial boundary, the ARs are reassigned to SC24 or SC25 and the 
corresponding compliance rates are recomputed. The resulting variation is included in the systematic uncertainty range of the compliance rates, together with the uncertainties caused by the artificial choice of the magnetic field threshold and the AR patch size as described in Section~\ref{S-DATA}. Table~\ref{tbl-FRACTION} summarizes the fractions of ARs that comply with the empirical laws for the full sample and for SC24 and SC25 separately. For both the Hale's and Joy's law, the compliance rate is 67.5\% in the full sample, 66.7\% in SC24, and 68.4\% in SC25; for the HHR, the compliance rate is 67.5\% in the full sample, 71.5\% in SC24, and 62.7\% in SC25. The fractions of the AH and AHJ classes are much smaller than those of the other classes. Overall, these compliance rates fall within the ranges reported by previous studies \cite[]{1998ApJLongcope, 2001ASPCPevtsov, 2012ApJLY, 2012ApJLi, 2018ApJ...867...89L, 2020ApJPark, 2021ApJ...920...31M}. 

\begin{figure} [ht]
\centerline{\includegraphics[width=1\textwidth,clip=]{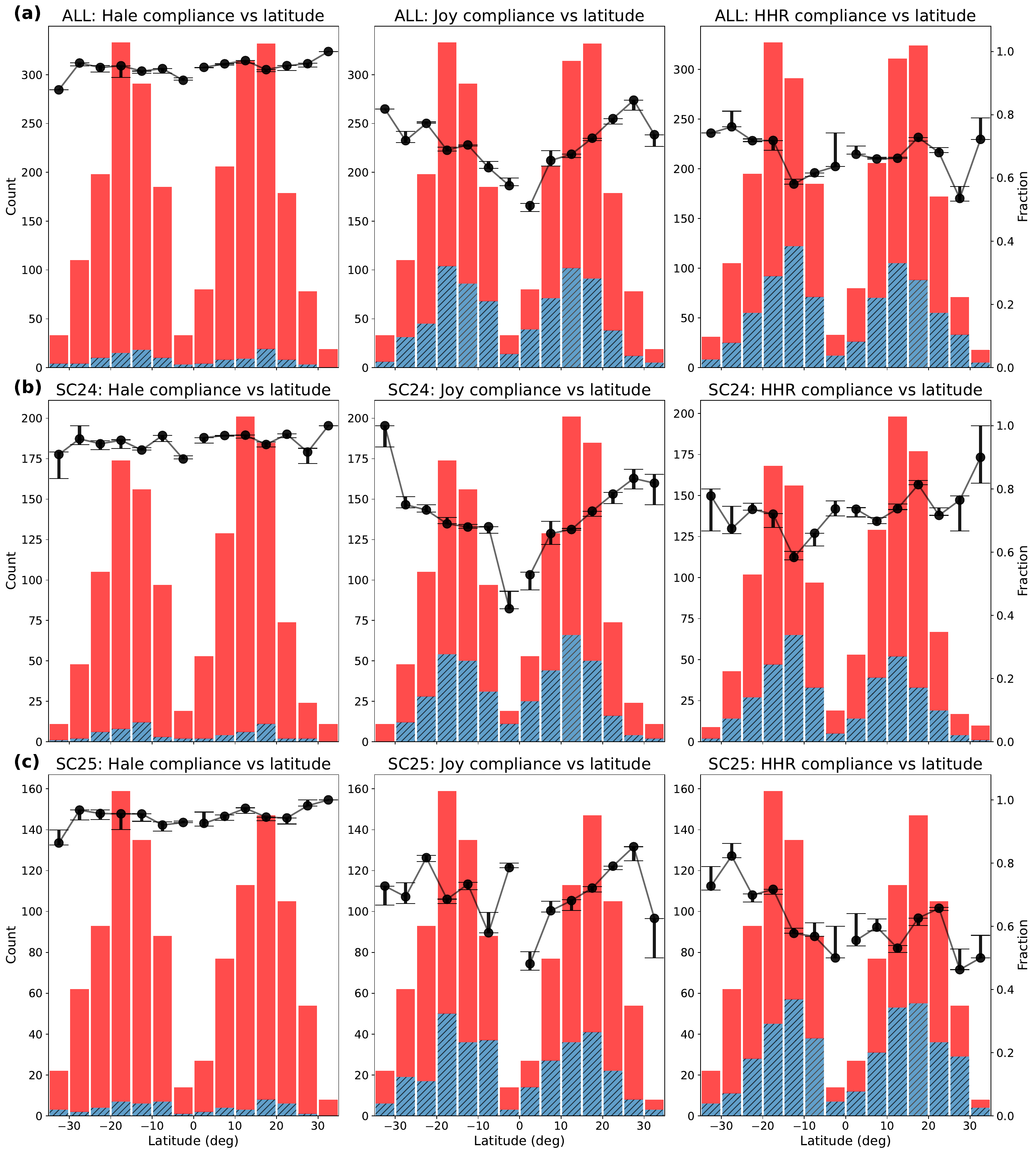}}
\small
        \caption{Frequency distribution of ARs with respect to latitude.
        The left, middle, and right column show the compliance state with Hale's law, Joy's law, and the HHR, respectively, for the full dataset (top row), SC24 (middle row), and SC25 (bottom row). The red (blue) bars, as scaled by the left $y$-axis, represent the number of ARs that obey (violate) the empirical laws; black dots, as scaled by the right $y$-axis, show the compliance rate in each individual latitude bins. ARs located at latitudes beyond $\pm30^\circ$ are grouped into the two tail bins. Error bars show the range of compliance rates obtained by varying the magnetic field threshold, the AR patch size, and the cycle boundary, as described in (Section~\ref{S-DATA}).}
\label{fig-COMLIANCEVSLATITUDE} 
\end{figure}

We then investigate the latitude dependence of compliance with the empirical laws. Figure~\ref{fig-COMLIANCEVSLATITUDE}(a-c) show the latitudinal distribution for the full sample, SC24, and SC25, respectively. The latitude ranges from -35$^\circ$ to 35$^\circ$, and is divided to 14 bins with 5$^\circ$ for each bin. ARs located at latitudes above 30$^\circ$ or below -30$^\circ$ are grouped into the two tail bins to compensate for the small sample size. The red bars represent ARs that obey the laws, while the blue bars represent those that violate the laws. The black lines show the fraction of ARs that obey the laws in each latitude bin. The error bars indicate the uncertainties estimated in the data reduction process (see Section~\ref{S-DATA}). The compliance rate varies only slightly with latitude for Hale's law, but more pronounced for Joy's law: it is generally higher at high latitudes than those at low latitudes. This trend is more evident in SC24 than in the ascending phase of SC25. For the HHR, 
the latitude dependence is less conclusive because of the larger fluctuations. 

The time dependence of the law compliance is shown in Figure~\ref{fig-COMLIANCEVSTIME} with the similar format as Figure~\ref{fig-COMLIANCEVSLATITUDE}, but it should be noted that the statistics of Hale's law and Joy's law extend back to 2008 December, whereas the HHR statistics start from 2010 May when HMI vector magnetograms are first available. There is also a tendency that the yearly compliance rate of ARs gradually decreases with time in each cycle. This implies that ARs in the ascending phase of a solar cycle are more likely to comply with these empirical laws than those in the declining phase. 

\begin{figure} [ht]
\centerline{\includegraphics[width=1\textwidth,clip=]{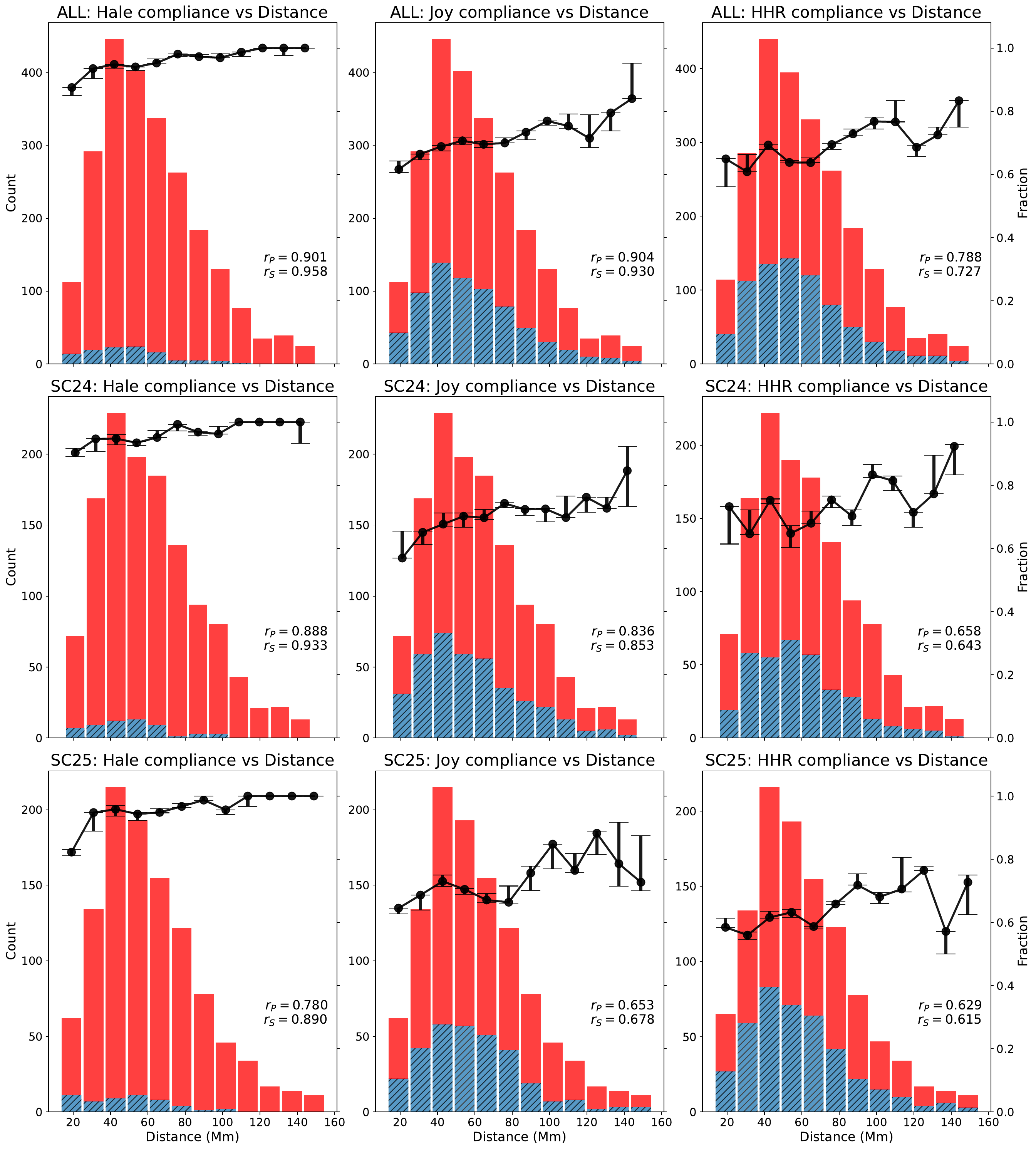}}
\small
        \caption{Frequency distribution of ARs with respect to centroid distance between opposite polarities. The left, middle, and right column show the compliance state with Hale's law, Joy's law, and the HHR, respectively, for the full dataset (top row), SC24 (middle row), and SC25 (bottom row). The red (blue) bars, as scaled by the left $y$-axis, represent the number of ARs that obey (violate) the empirical laws; black dots, as scaled by the right $y$-axis, show the compliance rate in each individual distance bins. The Pearson correlation coefficient ($r_{\mathrm{P}}$) and the Spearman rank correlation coefficient ($r_{\mathrm{S}}$) between compliance rate and centroid distance are annotated in each panel. Error bars show the range of compliance rates obtained by varying the magnetic field threshold, the AR patch size, and the cycle boundary, as described in (Section~\ref{S-DATA}).}
\label{fig-COMLIANCEVSDISTANCE} 
\end{figure}

\begin{figure}[ht]
\centerline{\includegraphics[width=1\textwidth,clip=]{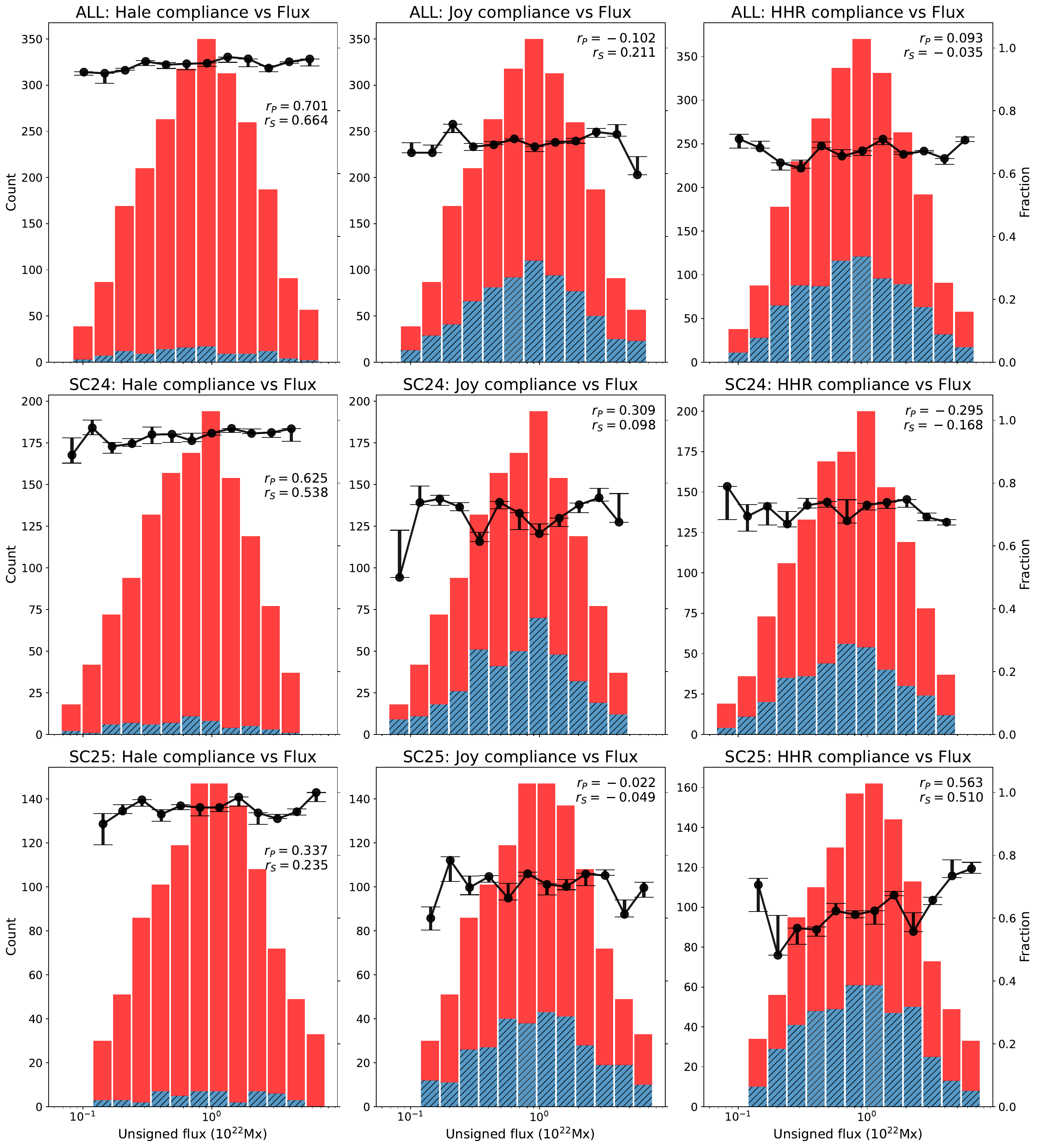}}
\small
        \caption{Frequency distribution of ARs with respect to total unsigned magnetic flux. The left, middle, and right column show the compliance state with Hale's law, Joy's law, and the HHR, respectively, for the full dataset (top row), SC24 (middle row), and SC25 (bottom row). The red (blue) bars, as scaled by the left $y$-axis, represent the number of ARs that obey (violate) the empirical laws; black dots, as scaled by the right $y$-axis, show the compliance rate in each individual flux bins. The Pearson correlation coefficient ($r_{\mathrm{P}}$) and the Spearman rank correlation coefficient ($r_{\mathrm{S}}$) between compliance rate and total unsigned flux are annotated in each panel. Error bars show the range of compliance rates obtained by varying the magnetic field threshold, the AR patch size, and the cycle boundary, as described in (Section~\ref{S-DATA})}.
\label{fig-COMLIANCEVSFLUX} 
\end{figure}

The compliance rate of ARs is further examined as the function of the centroid distance (Figure~\ref{fig-COMLIANCEVSDISTANCE}) and the total unsigned magnetic flux (Figure~\ref{fig-COMLIANCEVSFLUX}). In each panel, we calculate both the Pearson correlation coefficient ($r_{\mathrm{P}}$) and the Spearman rank correlation coefficient ($r_{\mathrm{S}}$) between the binned compliance rate and the corresponding AR properties. The former measures linear relationships between two variables, while the latter assesses their monotonic relationships by ranking the data. These two coefficients serve as descriptive measures of the binned trends. Similar signs and comparable magnitudes indicate a relatively robust monotonic trend that is close to linear, while large differences between them suggest that the dependence may be weaker or nonlinear. For the centroid distance, $r_{\mathrm{P}}$ and $r_{\mathrm{S}}$ are generally positive and comparable in all the panels of Figure~\ref{fig-COMLIANCEVSDISTANCE}, supporting a robust increase in compliance rate with increasing centroid distance. In Figure~\ref{fig-COMLIANCEVSFLUX}, the coefficients are generally weaker and less consistent. A clear positive dependence on unsigned flux is found for Hale's law, i.e., a tendency for ARs with higher unsigned flux to be more likely to comply with the Hale's law, but no clear dependence on unsigned flux is found for Joy's law nor the HHR.

\subsection{Statistical test on the relationship between the AR flare productivity and the empirical laws} \label{SS-FLARE}

To understand whether an AR's compliance with the empirical laws affects its flare productivity, we integrate all C-class and above flares associated with each individual AR from the Heliophysics Event Knowledgebase \cite[HEK]{2012SoPhHEK}, and then calculate the flare index as a proxy of the AR's flare productivity \cite[]{2010ApJJing, 2018SoPhLee, 2021SoPhChen}:
\begin{equation}
    \mathrm{FI}=\frac{1}{\tau}\sum (1 \times I_C+10 \times I_M + 100 \times I_X)
\end{equation}
where $\tau$ denotes the time range from the first flare to the last flare that occurred in the AR, which is measured in days, and $I_C,~I_M,~I_X$ represent the GOES peak soft X-ray flux of C-class, M-class, and X-class flares, respectively. In other words, FI measures the average daily flare productivity of a given AR.

We employ the Mann-Whitney U test \citep{corder2009nonparametric} to quantify the statistical significance of the difference in FIs between two AR groups. This nonparametric test assesses whether two independent samples of sizes $n_1$ and $n_2$ are drawn from the same distribution without normality assumption, and is therefore well suited to our FI data, which are highly skewed and contain a large fraction of zeros. By ranking all values from both samples jointly and comparing the sum of ranks between the two samples, the test intends to show whether the two samples are randomly mixed or clustered at opposite ends. In the former situation, there exists no significant difference between the ranks of these two groups of data, while the opposite is true for the latter situation \cite[]{2024ApJLiu, 2025ApJPan}. Let $\sum R_i$ be the sum of ranks for sample $i$ ($i=1,2$). The U statistics are
\begin{equation}
U_1 = n_1 n_2 + \frac{n_1(n_1+1)}{2} - \sum R_1, \qquad
U_2 = n_1 n_2 + \frac{n_2(n_2+1)}{2} - \sum R_2,
\end{equation}
The smaller of the two $U$ statistics is then used to compute the $z$-score under a normal approximation:
\begin{equation}
    z=\frac{U_i -\overline{x_U}}{S_U},
\end{equation}
where $\overline{x_U}=n_1n_2/2$ is the mean, and $S_U=\sqrt{n_1 n_2 (n_1 + n_2)/12}$ is the standard deviation. The null hypothesis that there is no tendency for ranks of the FIs of one group of ARs to be significantly different than those of the other. Hence, the Mann-Whitney test statistic is evaluated in a two-sided sense, so the significance depends only on the magnitude but not the sign of $z$. To facilitate interpretation of the direction of the difference, we assign $z$ the sign of the difference between the sum of the ranks, i.e., $\mathrm{sign}(\sum R_1-\sum R_2)$. With this convention, $z>0$ indicates that sample 1 tends to have larger ranks (and thus larger FI) than sample 2, whereas $z<0$ indicates the opposite. Here we use the significance level $\alpha$ to denote the probability of rejecting 
the null hypothesis when it is actually true, namely, the Type 1 error rate. At a significance level of $\alpha=0.05$, the critical range of $z$ score is $-1.96 \leq z \leq 1.96$ for a two-tailed test. In other words, if $z$ falls in this range, we cannot reject the null hypothesis. If the significance level is $\alpha=0.1$, the critical range of $z$ score is $-1.645 \leq z \leq 1.645$, and the result is stated to bear a marginal trend, indicating that the difference in FI between the two groups is weak \cite[]{statis01, Gelman01112006}. We further quantify the magnitude of the difference using an effect size (ES) based on the standardized statistic:
\begin{equation}
\mathrm{ES}=\frac{|z|}{\sqrt{n}},
\end{equation}
where $n=n_1+n_2$ is the total sample size. Following \cite{Cohen}, ES is categorized as small (0.1), medium (0.3), and large (0.5).

The test results are shown in Table~\ref{tbl-MWTEST}. For clarity and to keep the results concise, we perform pairwise Mann-Whitney U tests on three categories, i.e. the full sample, SC24, and SC25. For each category, the tests are implemented under three grouping schemes by considering (i) Hale's and Joy's laws only, (ii) the HHR only, and (iii) all three empirical laws simultaneously. In the HHR-only scheme, we denote ARs that satisfy the HHR as $\mathrm{HHR}|1$ and those that violate the HHR as $\mathrm{HHR}|0$. To avoid an excessive number of subgroups in the scheme considering three laws, we merge all ARs that violate either Hale's law or Joy's law into a single class, denoted as $\mathrm{ABN}$ for `abnormal', while ARs that satisfy both Hale's and Joy's laws are denoted as NOR for `normal'. We then indicate the compliance of HHR by appending a binary flag following `$|$': `$|$1' indicates HHR compliance and `$|$0' indicates HHR violation. For example, $\mathrm{NOR}|1$ denotes ARs that satisfy Hale's law, Joy's law, and the HHR simultaneously, whereas $\mathrm{ABN}|0$ denotes ARs that violate at least one of Hale's or Joy's laws and also violate the HHR. 

When considering all the empirical laws, i.e., in the test theme (NOR vs ABN)$|$HHR, only the pair $\mathrm{NOR}|0$ vs $\mathrm{NOR}|1$ shows statistical significance with $z=-2.333 < -1.96$ in the SC24 category. Thus we add one more pair $\mathrm{NOR}|0$ vs `Other' to each category, where ``Other'' denotes the combination of NOR$|$1, ABN$|$0, and ABN$|$1. In the SC24 sample, the test result of the pair $\mathrm{NOR}|0$ vs Other is also statistically significant, as indicated by a $z$-score of 2.273 exceeding the critical value. The results indicate that the ARs obeying Hale's law and Joy's law but violating HHR holds stronger flare productivity in SC24 than other types of ARs, while the results from the full sample and SC25 are not significant. When considering only HHR, the test results of SC24 show a marginal trend that ARs violating HHR (HHR$|$0) show a stronger flare productivity than ARs obeying HHR, while the results from the full sample and SC25 still show no statistical significance. Despite the statistical significance judged from $z$-scores, the correlation between the AR types and the flare productivity is quite weak, as indicated by the ES values of these test results (all less than 0.1). 

When focusing on Hale's law and Joy's law, we notice that ARs of AHJ type in SC24 show stronger flare productivity than the other three types of ARs, with $z=-2.449$ for AH vs AHJ, $z=-2.174$ for AJ vs AHJ, and $z=-2.073$ for NOR vs AHJ. Besides the results showing statistical significance for all these groups, the ES for the AH vs AHJ test is 0.306, indicating a medium strength correlation, while the ES for the AJ vs AHJ test is 0.109, representing a small-strength correlation according to Cohen's convention \cite[]{Cohen}. These results seem to suggest that the AHJ group has a relatively higher flare productivity compared to other groups, particularly when AHJ is tested against the AH and AJ groups. But the result is different for the full sample and SC25. There is a marginal trend that the flare productivity of AHJ is weaker than that of AJ and NOR in SC25. The distinct behavior of AHJ deserves further investigation, so we further analyze the magnetic complexity of the AHJ type ARs below.

NOAA SRS also provides the Mount Wilson classification of each AR shown by four parameters as a proxy of the magnetic complexity, i.e., $\alpha$ (unipolar), $\beta$ (bipolar), $\gamma$ (multipolar), and $\delta$. The parameter $\delta$ is assigned to regions where at least one sunspot has opposite magnetic polarities within a common penumbra \cite[]{1987SoPhZirin}. The $\delta$-ARs are generally regarded as having strong potential on producing flares. NOAA SRS provides the daily magnetic type of each AR, so an AR in our sample is regarded as a $\delta$-AR only if the $\delta$ configuration is recognized more than twice in the NOAA SRS records or it would be classified as being non-$\delta$. After classifying ARs into $\delta$ or non-$\delta$, we found that in the SC24 sample, among 36 AHJ ARs, 4 (11\%) are classified as $\delta$-ARs. In contrast, in the SC25 AHJ sample, only 1 (4\%) out of 24 ARs is classified as a $\delta$-AR. The $\delta$-ARs in SC24 include ARs 11429, 11990, 12158, and 12320, with FI of 83.2, 53.1, 21.2, and 8.7, respectively, while the only $\delta$-AR in SC25 is AR 13575 with FI of 66.5. Besides these ARs with strong flare productivity, almost all other AHJ type ARs in SC24 produce several C-class flares and thus their FI are not zero, while most of the AHJ type ARs in SC25 produce no flares above C-class. Hence, the AHJ type ARs' outstanding flare productivity in SC24 but weak performance in SC25 is significantly modulated by magnetic complexity. Further control variable analysis is necessary but not viable for such a small sample size.

\subsection{Dependence of the flare productivity on centroid distance and total unsigned flux}
\label{SS-FLAREONDISANDFLUX}

The dependence of the FI on centroid distance and total unsigned flux of ARs is demonstrated in Figure~\ref{fig-FIVSDISANDFLUX}. Panels (b) and (c) show the distribution of the median FI among 80 uniform bins. If more than half of the ARs in a bin have a zero FI, the median FI for that bin is zero. It is obvious that a jump exists at the first bin with non-zero median FI. Under this binning scheme, the threshold for the centroid distance and the unsigned flux is 50.5 Mm and $0.58\times10^{22}$ Mx, respectively, as indicated by the gray dashed lines in Figure~\ref{fig-FIVSDISANDFLUX}(b \& c). It is important to note that the threshold distinguishes bins with zero median FI from those with non-zero median FI. Different binning schemes give comparable results: varying the number of bins from 70 to 90 uniform bins yields a centroid distance threshold in the range of 48.7--51.3 Mm and an unsigned flux threshold in the range of 0.59--$0.64\times10^{22}$ Mx; with an equal-frequency binning scheme, in which each bin contains approximately the same amount of ARs, the thresholds become 50.2 Mm and 0.63 $\times10^{22}$ Mx, respectively.

Although the total unsigned flux and centroid distance are moderately correlated, with a Pearson correlation coefficient of 0.54 (Figure~\ref{fig-FIVSDISANDFLUX}a), the combination of their thresholds as obtained from Figure~\ref{fig-FIVSDISANDFLUX}(b \& c) effectively separates flare-productive ($\mathrm{FI}>0$) from flare-quiet ($\mathrm{FI}=0$) ARs: most flare-productive ARs are located in the right-half plane, especially in the top right quadrant. Using the threshold pairs derived from different binning schemes, we found that 60.9\% to 62.8\% of flare-productive ARs are selected by the thresholds. Conversely, 72.1\%--72.7\% of the ARs located above both thresholds are flare-productive.

\section{Discussion and Conclusion}
\label{S-discussion}

In this study, we have performed a comprehensive statistical analysis of ARs spanning SC24 and the ascending phase of SC25 (from 2008 December to 2025 December). Utilizing the deep learning-based SARD model, we automatically extracted ARs from SOHO/MDI and SDO/HMI LoS magnetograms and cross-matched them with NOAA SRS records. We investigated the compliance of these ARs with three empirical laws, Hale's law, Joy's law, and the HHR, and examined the relationship between the law compliance and flare productivity, which is quantified by the FI of each AR. The main findings are summarized and discussed as follows.

The majority of the ARs comply with the empirical laws. The overall compliance ratios for Hale's law and Joy's law is 67.5\%, while the compliance ratio of HHR is 67.3\% (Section \ref{S-DATA}, Table \ref{tbl-FRACTION}). These results are consistent with previous studies regarding earlier solar cycles \cite[]{1989SoPhWang, 1998ApJLongcope, 2001ASPCPevtsov, 2013ApJWYM, 2014ApJLiu, 2023SSRvNorton, 2025ApJQin}. We also investigated the dependencies of the compliance with these empirical laws on the physical properties of ARs. For centroid distance, the compliance rates of all three empirical laws show clear positive correlations with the binned distance, where both the Pearson coefficients $r_{\mathrm{P}}$ and the Spearman coefficients $r_{\mathrm{S}}$ remain positive in the full sample and the cycle subsets, indicating a robust increase in compliance rate with increasing centroid distance. For the total unsigned flux, the correlations are much weaker and less uniform. Hale's law still shows a positive correlation with unsigned flux ($r_{\mathrm{P}}=0.701$ and $r_{\mathrm{S}}=0.664$) for the full sample, whereas the corresponding coefficients for Joy's law and the HHR are generally small and, in some cases, change sign among different cycle subsets, suggesting that their dependence on unsigned flux is weak and not robust. Overall, ARs with larger centroid distance are more likely to comply with the empirical laws, while a positive dependence on total unsigned flux is evident mainly for Hale's law (Section~\ref{SS-DISTRIBUTION2425}). Further, the compliance rate tends to increase with latitude and is higher during the ascending phase of the solar cycle compared to the declining phase. These findings are generally consistent with previous studies \cite[]{1989SoPhWang, 1998ApJLongcope, 2005PASJHagino, 2010A&ADasi, Stenflo&Kosovichev2012, 2013SoPhWeber, 2020ApJPark}.

The dependence of law compliance on AR magnetic flux and centroid distance has been the subject of several prior investigations \cite[]{1989SoPhWang, 1999SoPhTian,Stenflo&Kosovichev2012}. More recently, \cite{2018ApJLijing} used magnetograms from SDO/HMI and SOHO/MDI spanning SC23 and SC24 to investigate the differences and similarities between Hale and anti-Hale sunspots, and found that the average total unsigned flux and average centroid distance of Hale spots are larger than that of anti-Hale spots. The author attributed the results to the combined effect of magnetic buoyancy, magnetic tension, and Coriolis force within these flux systems, as indicated by \cite{1993A&ADsilva}. \cite{2024ApJSreedevi} used the magnetograms of 12173 unique bipolar magnetic regions from 1996 to 2023 in the AutoTAB catalog \cite[]{2023ApJSSreedevi} derived from SOHO/MDI and SDO/HMI, and found that bipolar magnetic regions with higher magnetic flux and larger centroid distance exhibit better compliance with Joy’s law. \cite{2020ApJPark} estimated magnetic helicity fluxes in 1105 ARs from 2010 to 2017 to examine the HHR, and found that ARs appearing at higher latitudes during the earlier inclining phase of the solar cycle and having larger total unsigned flux tend to obey HHR. Our findings that ARs with larger magnetic flux and larger centroid distance tend to comply with the empirical laws align well with these previous studies, supporting a unified physical mechanism that small, weak flux tubes are more susceptible to the buffeting effects of turbulent convection during their ascent through the convection zone, which can randomize their tilt angles and twist, leading to violations of Joy's law and the HHR. In contrast, widely separated and high-flux ARs possess sufficient magnetic tension to resist convective disruption, allowing the Coriolis force to effectively imprint the systematic tilt and twist as indicated by \cite{1993A&ADsilva}.

A key objective of this work was to investigate the relationship between the compliance of the empirical laws and the flare productivity of ARs (Section \ref{SS-FLARE}, Tabel~\ref{tbl-MWTEST}). To quantify this relationship, we utilized the FI as a proxy for AR flare productivity. We employed the non-parametric Mann-Whitney U test to conduct pairwise comparisons between different AR groups that are classified according to compliance with the empirical laws. While the analysis of the full sample revealed no statistically significant differences among the groups, ARs in SC24 that comply with both Hale's and Joy's laws but violate the HHR exhibited significantly higher flare productivity compared to other categories (Table \ref{tbl-MWTEST}). This scenario is reminiscent of the flare trigger model proposed by \cite{2003AdSpRKusano}, where the injection of reverse magnetic helicity would trigger magnetic reconnection and caused solar flares. This model is supported by several observations that the annihilation of opposite-signed magnetic helicity is associated with the onset of solar flares \cite[]{2010ApJ...720.1102P, 2013ApJPark, 2012ApJ...761...60V}. In the context of our findings, ARs that obey Hale's and Joy's law but violate the HHR represent emerging flux systems with a standard bipolar orientation but an anomalous internal twist. The violation of the HHR implies that these regions carry magnetic helicity opposite to the dominant hemispheric sign. Consequently, as these ARs interact with the pre-existing coronal magnetic field, which typically follows the hemispheric sign, they naturally create a topology favorable for the interaction and annihilation of opposite-signed magnetic helicity, thereby enhancing flare productivity. 

The statistical analysis reveals a disparity regarding the AHJ type that violates both Hale's and Joy's law (Section \ref{SS-FLARE}, Table \ref{tbl-MWTEST}). In SC24, the AHJ-type ARs exhibit significantly higher flare productivity than other types (NOR, AH, and AJ), while the same type of ARs in SC25 shows a marginal trend of lower flare productivity than those of NOR and AJ types. Detailed inspection reveals that the SC24 AHJ sample contains a higher proportion of flare-productive $\delta$-type ARs and maintains a non-zero baseline of C-class flaring activities, in contrast to the largely quiescent SC25 AHJ sample. Given the rank-based nature of the Mann-Whitney U test and the small sample size, we tentatively attribute the elevated flare productivity of the AHJ type in SC24 to the statistical bias introduced by the uncontrolled variable of magnetic complexity, rather than its intrinsic property.

\begin{figure}[ht]
\centerline{\includegraphics[width=1\textwidth,clip=]{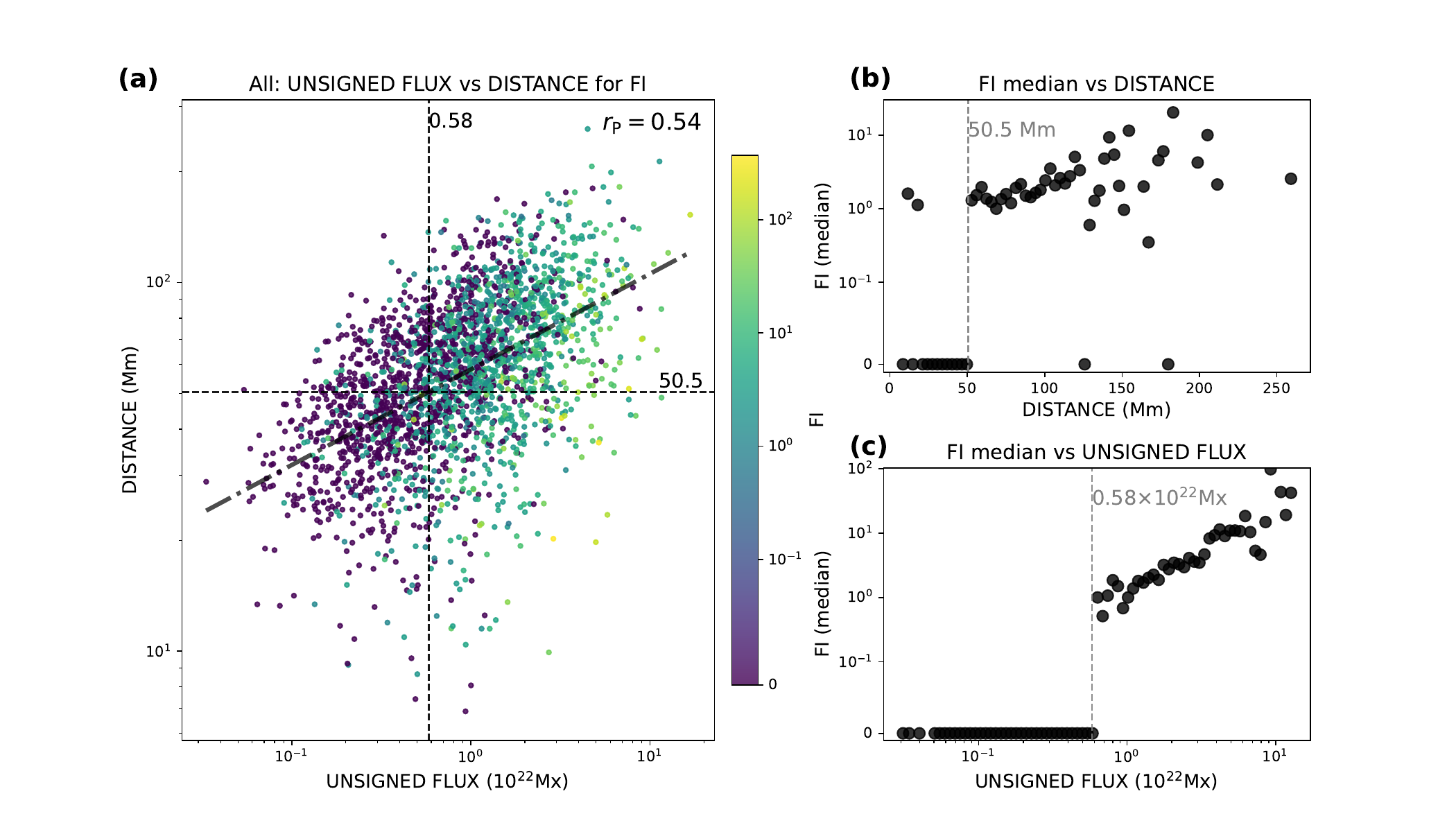}}
\caption{\small The distribution of FI with respect to centroid distance and unsigned flux. (a) Scatter plot of unsigned flux versus centroid distance for all ARs, color-coded by FI. The vertical and horizontal dashed lines mark the flux and distance thresholds identified from the binned median FI distributions in panels (b) and (c). The dot-dashed line indicates the linear fitting between unsigned flux and centroid distance while the Pearson correlation coefficient ($r_{\mathrm{P}}$) is shown in the upper right corner. The distribution of median FI with respect to centroid distance (b) and  unsigned flux (c), respectively, is computed within each individual bins; the dashed line indicates the distance and flux threshold at which the binned median FI shows a jump. \label{fig-FIVSDISANDFLUX} } 
\end{figure}

Finally, The investigation on the FI dependence on AR properties reveals an empirical threshold of 50.5 Mm for centroid distance and of 0.58 $\times10^{22}$ Mx for total unsigned magnetic flux (Section \ref{SS-FLAREONDISANDFLUX}, Figure \ref{fig-FIVSDISANDFLUX}). Below these values, the median FI of binned ARs drops to zero, suggesting that a minimal spacial scale and magnetic energy are required for significant flaring activities, regardless of whether the ARs comply with the empirical laws. Although small and weak ARs are susceptible to convective perturbations and therefore prone to violate the empirical laws, they typically lack sufficient magnetic free energy to power flares.

To conclude, the majority of ARs comply with Hale's law, Joy's law, and the HHR. The compliance rate tends to be higher for ARs emerging at higher latitudes, i.e., during the ascending phase of a solar cycle, and for those having larger centroid distance and stronger unsigned flux. However, the compliance with these empirical laws alone does not provide a reliable indicator for the flare productivity of ARs, but mainly reflect the statistical properties of emerging flux systems. Instead, flaring activity is more closely associated with sufficiently large and strong magnetic systems.

\begin{acknowledgments}
This work was supported by the National Key R\&D Program of China (2022YFF0503002), the Strategic Priority Program of the Chinese Academy of Sciences (XDB0560102), and the NSFC (42274204, 12373064, 42188101, 11925302, 12425305). The authors thank the Joint Science Operations Center (JSOC) at Stanford University for the LoS magnetograms of HMI and MDI used in this study.
\end{acknowledgments}

\bibliography{sample7}{}
\bibliographystyle{aasjournalv7}

\begin{deluxetable}{lcccccc}[ht]
\tabletypesize{\normalsize}
\tablewidth{\textwidth}
\tablecaption{Compliance rates of empirical laws for all the data, SC24, and SC25 data set, respectively. \label{tbl-FRACTION}}
\tablehead{
\colhead{Sample} & \colhead{NOR\tablenotemark{a}} & \colhead{AH\tablenotemark{b}} & \colhead{AJ\tablenotemark{c}} & \colhead{AHJ\tablenotemark{d}} & \colhead{HHR$|$1\tablenotemark{e}} & \colhead{HHR$|$0\tablenotemark{f}}
}
\startdata
All  & 67.5\% & 2.3\% & 27.7\% & 2.5\% & 67.3\% & 32.7\% \\
SC24 & 66.7\% & 2.0\% & 28.5\% & 2.8\% & 71.5\% & 28.5\% \\
SC25 & 68.4\% & 2.7\% & 26.7\% & 2.2\% & 62.7\% & 37.3\% \\
\enddata
\tablenotetext{a}{ARs obeying both Hale's law and Joy's law}
\tablenotetext{b}{ARs violating Hale's law but obeying Joy's law}
\tablenotetext{c}{ARs violating Joy's law but obeying Hale's law}
\tablenotetext{d}{ARs violating both Hale's law and Joy's law}
\tablenotetext{e}{ARs obeying HHR}
\tablenotetext{f}{ARs violating HHR}
\end{deluxetable}

\begin{deluxetable}{lcccrrrr}
\tablecaption{Mann-Whitney U test results on FI for different groups.\label{tbl-MWTEST}}
\tabletypesize{\scriptsize}
\tablewidth{\textwidth}
\tablehead{
\colhead{Dataset} & \colhead{Test} & \colhead{Group$_1$} & \colhead{Group$_2$} &
\colhead{$n_1$} & \colhead{$n_2$} & \colhead{$z$} & \colhead{ES}
}
\startdata
\tableline
ALL  & Hale-Joy-only & AH  & AJ  &   59 & 664 & -1.392 & 0.052 \\
     &              & NOR  & AH  & 1632 &  59 &  1.370 & 0.033 \\
     &              & AH  & AHJ &   59 &  60 & -1.194 & 0.109 \\
     &              & NOR  & AJ  & 1632 & 664 & -0.339 & 0.007 \\
     &              & NOR  & AHJ & 1632 &  60 & -0.271 & 0.007 \\
     &              & AJ  & AHJ &  664 &  60 & -0.116 & 0.004 \\
\tableline
ALL  & HHR-only      & HHR$|$1 & HHR$|$0 & 1629 & 786 & -0.808 & 0.016 \\
\tableline
ALL  & (NOR vs ABN)$|$HHR & NOR$|$0   & Other    &  585 & 1830 &  1.322 & 0.027 \\
     &                   & NOR$|$1   & NOR$|$0    & 1047 &  585 & -1.465 & 0.036 \\
     &                   & NOR$|$0   & ABN$|$0  &  585 &  201 &  1.117 & 0.040 \\
     &                   & NOR$|$1   & ABN$|$1  & 1047 &  582 & -0.892 & 0.022 \\
     &                   & ABN$|$1 & ABN$|$0  &  582 &  201 &  0.750 & 0.027 \\
     &                   & NOR$|$0   & ABN$|$1  &  585 &  582 &  0.502 & 0.015 \\
     &                   & NOR$|$1   & ABN$|$0  & 1047 &  201 &  0.268 & 0.008 \\
\tableline
SC24 & Hale-Joy-only & AH  & AHJ &  28 &  36 & -2.449 & 0.306 \\
     &              & AJ  & AHJ & 361 &  36 & -2.174 & 0.109 \\
     &              & NOR  & AHJ & 849 &  36 & -2.073 & 0.070 \\
     &              & NOR   & AH  & 849 &  28 &  1.469 & 0.050 \\
     &              & AH  & AJ  &  28 & 361 & -1.249 & 0.063 \\
     &              & NOR   & AJ  & 849 & 361 &  0.546 & 0.016 \\
\tableline
SC24 & HHR-only      & HHR$|$1 & HHR$|$0 & 915 & 359 & -1.736 & 0.049 \\
\tableline
SC24 & (NOR vs ABN)$|$HHR & NOR$|$0   & Other    & 257 & 1017 &  2.273 & 0.064 \\
     &                   & NOR$|$1   & NOR$|$0    & 592 &  257 & -2.333 & 0.080 \\
     &                   & NOR$|$0   & ABN$|$1  & 257 &  323 &  1.492 & 0.062 \\
     &                   & NOR$|$0   & ABN$|$0  & 257 &  102 &  1.439 & 0.076 \\
     &                   & NOR$|$1   & ABN$|$1  & 592 &  323 & -0.659 & 0.022 \\
     &                   & ABN$|$1 & ABN$|$0  & 323 &  102 &  0.357 & 0.017 \\
     &                   & NOR$|$1   & ABN$|$0  & 592 &  102 &  0.026 & 0.001 \\
\tableline
SC25 & Hale-Joy-only & AJ  & AHJ & 303 &  24 &  1.866 & 0.103 \\
     &              & NOR   & AHJ & 783 &  24 &  1.715 & 0.060 \\
     &              & NOR   & AJ  & 783 & 303 & -1.004 & 0.030 \\
     &              & AH  & AHJ &  31 &  24 &  1.000 & 0.135 \\
     &              & AH  & AJ  &  31 & 303 & -0.793 & 0.043 \\
     &              & NOR   & AH  & 783 &  31 &  0.547 & 0.019 \\
\tableline
SC25 & HHR-only      & HHR$|$1 & HHR$|$0 & 714 & 427 &  0.832 & 0.025 \\
\tableline
SC25 & (NOR vs ABN)$|$HHR & NOR$|$0   & Other    & 328 & 813 & -0.580 & 0.017 \\
     &                   & NOR$|$0   & ABN$|$1  & 328 & 259 & -0.837 & 0.035 \\
     &                   & ABN$|$1 & ABN$|$0  & 259 &  99 &  0.695 & 0.037 \\
     &                   & NOR$|$1   & ABN$|$0  & 455 &  99 &  0.492 & 0.021 \\
     &                   & NOR$|$1   & ABN$|$1  & 455 & 259 & -0.482 & 0.018 \\
     &                   & NOR$|$1   & NOR$|$0    & 455 & 328 &  0.465 & 0.017 \\
     &                   & NOR$|$0   & ABN$|$0  & 328 &  99 &  0.231 & 0.011 \\
\tableline
\enddata
\tablenotetext{}{The ``ABN'' category integrates ARs classified as AH, AJ, or AHJ. The ``Other'' category includes NOR$|$1, ABN$|$0, and ABN$|$1.}
\end{deluxetable}

\end{document}